# Deep Learning Enables Reduced Gadolinium Dose for Contrast-Enhanced Blood-Brain Barrier Opening

Pin-Yu Lee, Hong-Jian Wei, Antonios N. Pouliopoulos, Britney T. Forsyth, Yanting Yang, Chenghao Zhang, Andrew F. Laine, Elisa E. Konofagou, Cheng-Chia Wu, and Jia Guo

***Abstract*—Focused ultrasound (FUS) can be used to open the blood-brain barrier (BBB), and MRI with contrast agents can detect that opening. However, repeated use of gadolinium-based contrast agents (GBCAs) presents safety concerns to patients. This study is the first to propose the idea of modeling a volume transfer constant (Ktrans) through deep learning to reduce the dosage of contrast agents. The goal of the study is not only to reconstruct artificial intelligence (AI) derived Ktrans images but to also enhance the intensity with low dosage contrast agent T1 weighted MRI scans. We successfully validated this idea through a previous state-of-the-art temporal network algorithm, which focused on extracting time domain features at the voxel level. Then we used a Spatiotemporal Network (ST-Net), composed of a spatiotemporal convolutional neural network (CNN)-based deep learning architecture with the addition of a three-dimensional CNN encoder, to improve the model performance. We tested the ST-Net model on ten datasets of FUS-induced BBB-openings aquired from different sides of the mouse brain. ST-Net successfully detected and enhanced BBB-opening signals without sacrificing spatial domain information. ST-Net was shown to be a promising method of reducing the need of contrast agents for modeling BBB-opening K-trans maps from time-series Dynamic Contrast-Enhanced Magnetic Resonance Imaging (DCE-MRI) scans.

## I. INTRODUCTION

SYSTEMIC therapy options for central nervous system (CNS) diseases, including brain tumors, Alzheimer's, or Parkinson's disease, have been limited due to the presence of the blood-brain barrier (BBB) [1], [11]-[13]. The BBB is a unique vascular feature of the CNS [2], [15]. Tight junctions connect adjacent cerebral endothelial cells to the highly regulated transport system of the endothelial cell membrane. Together they form a physiological diffusion barrier that maintains the homeostasis of the brain by protecting it from exogenous and endogenous substances, but also hinders the delivery of therapeutic agents to the brain [14]. Since the BBB prevents over 98% of small-molecule drugs and approximately 100% of large-molecule drugs from entering the brain parenchyma, this makes it a major limiting factor for systemic treatment of CNS diseases [12], [13]. Extensive and ongoing research has been done to optimize drug delivery by overcoming the BBB, including intracranial injections, hyperosmotic solutions, convection-enhanced delivery (CED), and focused ultrasound (FUS) [15], [16].

Several studies have shown that FUS with intravenously injected microbubbles can temporarily induce BBB-openings for non-invasive drug delivery to the brain [2]. FUS treatments are performed using acoustic waves similar to diagnostic ultrasound. However, instead of constructing images from echoes generated by tissue interfaces, FUS uses a transducer that concentrates acoustic waves at a focal point, where the acoustic energy is significantly augmented. This allows the sonication to generate mechanical effects, thermal effects, or both. Optimization of acoustic parameters and MB dosage has been shown to achieve local and reversible BBB-openings without damaging the brain parenchyma in multiple preclinical models, including rodents and non-human primates [17], [18]. Clinical advancement of FUS technology has progressed rapidly over the past few years with clinical trials showing safety with BBB-openings in both patients with brain tumors, amyotrophic lateral sclerosis, and Alzheimer's disease [19]-

Manuscript received January xx, 2023; revised January xx, 2023; accepted January xx, 2023. Date of publication January xx, 2023; date of current version xx xx, 2023. This work was supported in part by the Gary and Yael Fegel Family Foundation, St. Baldrick's Foundation, the Star and Storm Foundation, the Matheson Foundation (UR010590), Swim Across America, a Herbert Irving Cancer Center Support Grant (P30CA013696), Sebastian Strong Foundation, and the ZI Seed Grant for MR Studies Program.

P. Lee, B. T. Forsyth, Y. Yang, C. Zhang are with the Department of Biomedical Engineering, The Fu Foundation of Engineering and Applied Science, Columbia University, New York, NY 10027 USA (e-mail: pl2744@columbia.edu; btf2115@columbia.edu; yy3189@columbia.edu; cz2639@columbia.edu).
H. Wei and C. Wu are with the Department of Radiation Oncology, Columbia University Irving Medical Center, New York, NY 10032 USA (e-mail: hw2711@cumc.columbia.edu; cw2666@cumc.columbia.edu).
A. N. Pouliopoulos was with the Department of Biomedical Engineering, Columbia University. He is now with the School of Biomedical Engineering & Imaging Sciences, King's College London, London, UK (email: antonios.pouliopoulos@kcl.ac.uk)
A. F. Laine and E. E. Konofagou are with the Departments of Biomedical Engineering and Radiology (Physics), Columbia University, New York, NY 10027 USA (e-mail: al418@columbia.edu; ek2191@columbia.edu).
J. Guo is with the Department of Psychiatry, Columbia University Irving Medical Center, New York, NY 10032 USA (e-mail: jg3400@columbia.edu).

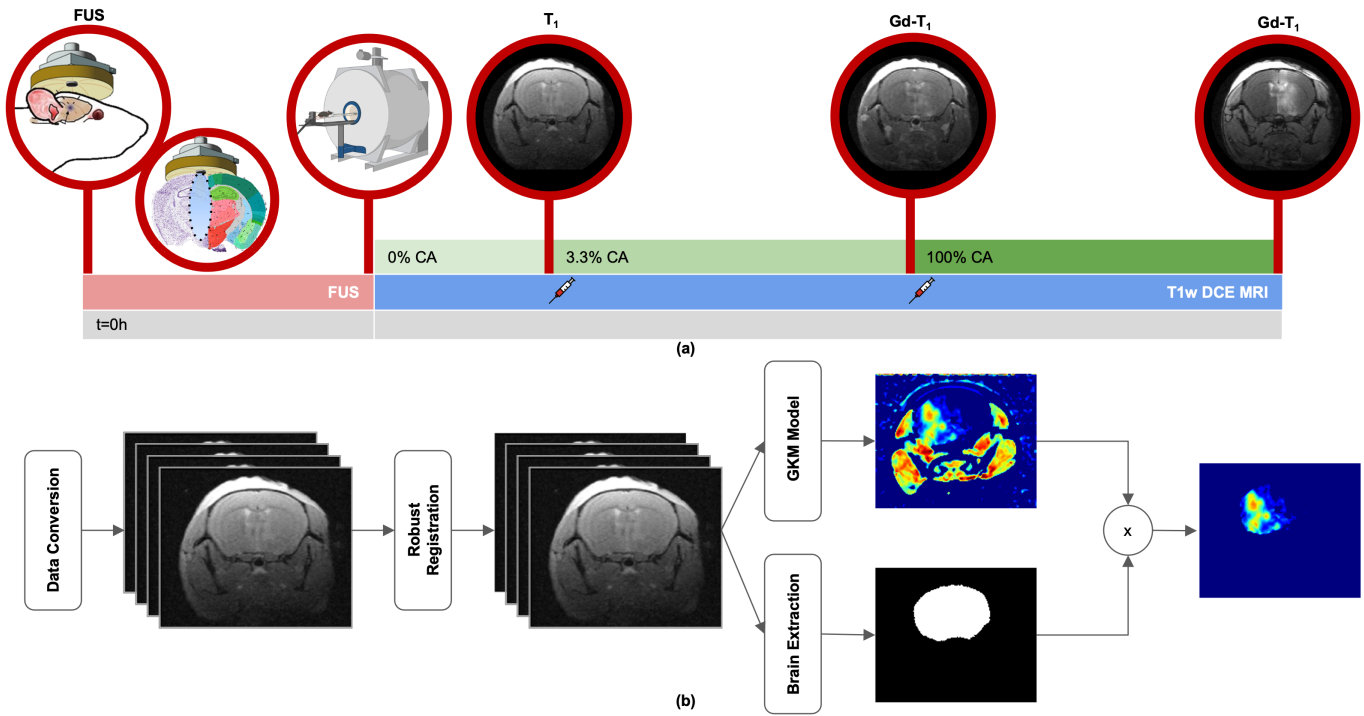

Fig. 1. (A) Timeline of the experimental procedure. Focused ultrasound (FUS) disrupts the blood-brain barrier (BBB) with the injection of microbubbles. After twelve hours, the BBB-opening was stable, and the mouse was placed on a Magnetic Resonance Imaging (MRI) system and scanned for the baseline for the first four acquisitions. We first injected 10 mmol/kg contrast agent (3.3% of the full dosage) gadodiamide and acquired eighty T1-weighted (T1W) dynamic contrast-enhanced (DCE)-MRI. Following the injection of the remaining 97.7% contrast agent (full dose), eighty-four T1W DCE-MRI were acquired. (B) Image preprocessing pipeline. We first converted the raw DCE-MRI images from DICOM format to NIFTI format and performed within-subject robust registration. We then generated the volume transfer constant (Ktrans) map through the general kinetic model (GKM) model. Finally, we extracted the whole brain Ktrans map with the manually labeled brain mask.

[22]. Multiple clinical trials are currently being conducted to study the safety and feasibility of BBB-opening. These trials may revolutionize drug-based therapy for intracranial disease.

Many strategies have been developed to identify the BBB-opening [23]. Among those strategies, FUS-induced BBB-opening is typically detected using Magnetic Resonance Imaging (MRI) with gadolinium-based contrast agents (GBCAs) using a T1 weighted sequence. However, there are limitations to this strategy. For drug delivery, systemic therapies can be given from daily to weekly to monthly, and repeated BBB-opening would theoretically be needed to ensure optimal drug delivery. To validate a BBB-opening, multiple MRI scans with contrast agents would be needed. It has been documented that repeated use of GBCAs can accumulate and be retained in body tissues, including the brain. This raises serious concerns for patient safety as renal impairment, specifically nephrogenic systemic fibrosis, may occur [4]-[7], [24]. In 2018, the FDA warned that gadolinium is retained in the organs after scanning by GBCA MRI, which has potential risks in humans based on toxicities observed in preclinical studies. Secondly, routine non-contrast scans are obtained in addition to post-contrast MRI scans. The additional contrast-based sequences can extend MRI scanning time. This can lead to increased costs, patient discomfort, and movement/motion artifacts. To address the potential safety concerns, it is necessary to develop alternative imaging techniques with reduced or no GBCAs.

Recently, a deep learning algorithm has been shown to extract diagnostic quality images with a 10-fold lower gadolinium dose than typically used, suggesting its potential to reduce GBCAs dose in brain MRI [25]. Further, using a deep contrast algorithm, artificial intelligence (AI) was able to predict regions of contrast enhancement in the absence of a contrast based on pixelated changes not observable to the human eye. Thus, we hypothesize that with deep-contrast AI, we can generate a computer algorithm to predict full-dose GBCAs BBB-openings using low-dose GBCAs T1 sequences. Furthermore, these low-contrast scanning sequences can be optimized to minimize overall scan time. Here, we present a proof-of-concept study in mice to demonstrate feasibility before conducting human trials.

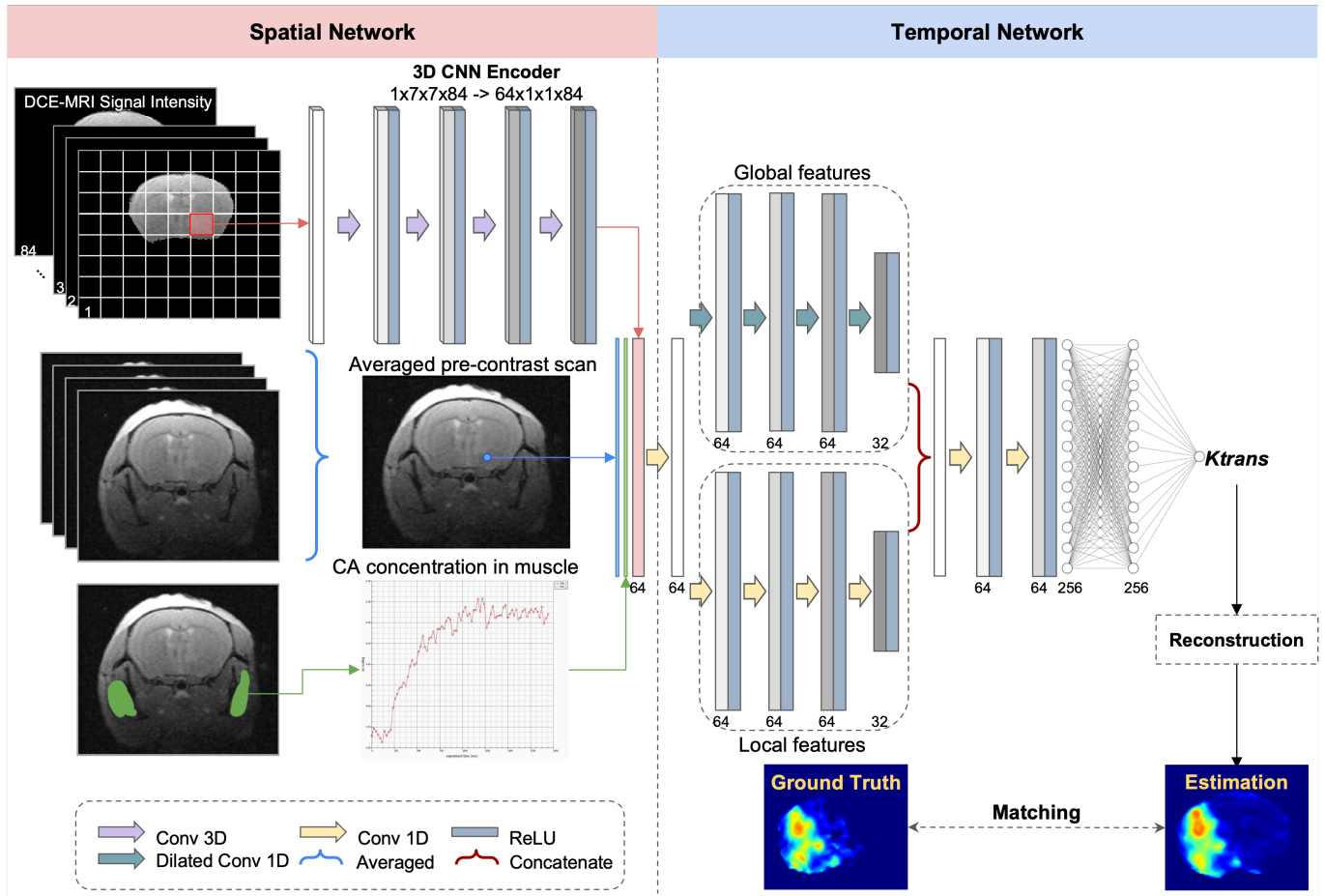

Fig. 2. Proposed ST-Net architecture. The four-dimensional dynamic contrast-enhanced (DCE)-Magnetic Resonance Imaging (MRI) scans were first cropped to 7x7x48 patches. We then extracted spatial information, using a three-dimensional convolutional neural network (CNN) encoder. Concatenation of spatial features with two reference arrays: (1) The average of DCE-MRI signal before contrast agent injection for each voxel (2) contrast agent concentration in muscle tissues. The size of the output features for each layer was provided in the figure. Each fully connected layer is followed by a Leaky ReLU activation. The output from the proposed ST-Net is a volume transfer constant (Ktrans) value, which was reconstructed to acquire a whole-brain Ktrans map

## II. METHODS

### A. FUS-induced BBB-opening Protocol

BBB-opening induced in murine brains by FUS with the administration of microbubbles has been described in detail [3]. Briefly, a single-element, spherical-segment FUS transducer was driven by a function generator through a 50-dB power amplifier. A single-element, pulse-echo transducer was housed within the central core of the FUS transducer and used for passive cavitation detection (PCD) of acoustic emissions. In-house manufactured microbubbles (concentration: $8×10^8$ bubbles /mL, diameter: $1.37 ± 1.02$ μm) were diluted in saline to 200 μL and injected intravenously. Sonication was delivered at 0.5 MHz with a peak-negative pressure of 0.3 MPa in bursts of 10 ms length at 5 Hz repetition time over 120 s (600 pulses).

### B. DCE-MRI Protocol and Data Acquisitions

Following the FUS procedure, the mouse was transferred to the Bruker BioSpec 94/20 scanner (field strength, 9.4 T; bore size, 30 cm) horizontal small animal MRI scanner with software ParaVision 6.0.1 (Bruker BioSpin, Billerica, MA, USA), an 86-mm inner diameter birdcage 1H volume transmits coil and a 1H mouse-head-only Cryogenic RF coil (CryoProbeTM). Mice were anesthetized using medical air and isoflurane (3% volume for induction, 1.1-1.5% for maintenance at 1 liter/min airflow, via a nose cone). The DCE-MRI images were acquired using a 2-D FLASH T1- weighted sequence ($180 × 150 × 18 × 84$ matrix size, spatial resolution of $100 × 100$ $μm^2$, slice thickness of 500 μm, TR/TE = 200/2.12 ms) before (i.e., the first four scans) and during the intraperitoneal (IP) injections of the contrast agent Gadodiamide (Gd) (Omniscan; GE Healthcare, Princeton, NJ, USA).

A contrast agent was used as a tracer to depict the area of the BBB-opening. We injected the contrast agent at two time points to obtain MRI scans with different volumes of contrast agent. We first injected 10 mmol/kg GBCAs, which is 3.3% of the full dosage of the GBCAs (low dose), then administered the remaining 97.7% GBCAs (full dose). For each of the injections, we acquired four-dimensional DCE T1-weighted anatomical brain MRI images with 18 slices and 48 acquisitions respectively. The total acquisition time for DCE-MRI was approximately 1 hour. The timeline for FUS and DCE-MRI image acquisition is shown in Fig. 1(a). Schematic showing the timeline for MRI image acquisition and image processing pipeline.

### C. Image Preprocessing

The raw DCE-MRI images were first converted from DICOM to NIFTI format and within-subject robust registration was performed using the FreeSurfer tool [42].

We used the volume transfer constant (Ktrans) generated by a MATLAB program as the desired ground truth for deep learning [9]. Ktrans denotes the transfer rate from the blood plasma to the extravascular extracellular space of each voxel, which is a voxel-level mapping quantified from the DCE protocol and models the capillary permeability. Therefore, it can be used to detect BBB-opening [34], [35]. The Ktrans map was calculated with two kinetic models, the general kinetic model (GKM) [9], [10], [26], [43] and the reference region model (RRM), to quantify the permeability [9].

In addition, we manually labeled the brain mask with 3DSlicer to extract and train the model with only whole brain (WB) information. The preprocessing pipeline is shown in Fig. 1(b).

### D. Deep Learning Model

#### 1) Spatial Network

Combining spatial and temporal deep learning networks, we designed an ST-Net, a voxel-based model to predict full dose Gd BBB-opening from low dose Gd DCE-MRI images. We first cropped each voxel of the WB scan over time to 7x7x48 patches and used a three-dimensional convolutional neural network (CNN) encoder to extract and preserve spatial features.

#### 2) Temporal Network

Following the spatial network, we concatenated the output one-dimensional array (64x84) with the other two channels, which were used as reference: (1) Broadcast a single value, the average of the four pre-contrast images, to the same length of frames (48 acquisitions), and (2) Averaged Gd concentration changes in muscle tissue area with time. The concatenated array (66x84) was conducted in a temporal network in reference to a previous model, fast-eTofts, proposed by Fang et al [8]. We first performed a one-dimensional CNN layer in the temporal model to fuse the spatial information with reference information and extract low-level temporal features. The following two parallel CNN pathways were used to extract long-term (global pathway) and short-term (local pathway) features. Finally, two one-dimensional CNN layers and a fully connected layer were used to fuse the long-term and short-term information and to predict the full dose Ktrans value for each pixel. Additionally, we added dropout layers after the fully connected layers to prevent model overfitting. We reconstructed the resulting Ktrans values to acquire a 3d WB Ktrans map. The ST-Net architecture is illustrated in Fig. 2.

#### 3) Model Hyperparameters

The ST-Net was trained using the Adam optimizer [31] and the loss function was defined as the mean absolute error (MAE) with early stopping at 300 epochs. We trained the ST-Net with several hyperparameters. To fine-tune the model, we set the network with batch size 512, learning rate 1e-4, and added four layers of a CNN encoder without batch normalization. All the models were trained on three 24 GB NVIDIA Quadro 6000 graphical processing units using PyTorch.

### E. Dataset Details

#### 1) Training and Testing Dataset Selection

We repeatedly selected two mice for testing, and the rest of the eight mice for training set data. The WB voxel of the eight mice was shuffled and split at the ratio of four to one for training and validation, respectively. The cross-validation strategy showed the robustness of the deep learning model.

#### 2) Strategies for Removal of Abnormal Values

Both the input DCE-MRI data and the ground truth Ktrans map did not apply any filters. All input data for the model (DCE-MRI patches, averaged pre-contrast scan, and Averaged Gd concentration in muscle) were normalized by the 99 percentiles of the averaged pre-contrast scan. The derived ground truth Ktrans maps had some extremely high values due to the noise. Therefore, to minimize the effect of the GKM model reconstruction errors, only voxels with the Ktrans value at the range of [0, 0.05] 1/min were considered when calculating the loss.

#### 3) BBB-Opening Patches Steps

The ST-Net tended to overfit due to the high overlap between BBB-opening patches in the training dataset. Therefore, we set a patch step rate to reduce the overlapping area between input patches.

#### 4) Using Two Regions of Interest as Input

The input voxels for the deep learning model were composed of two regions of interest (ROI) within the WB. We manually depicted the brain area and the BBB- opening region using 3DSlicer. For all the mice datasets, we chose two multiple slices of ROIs, with one encompassing all the voxels in BBB-opening, and the other from normal-appearing brain tissue having four-fold more voxels than BBB-opening ROI.

### F. Statistical Analysis

To evaluate the quality of the predicted K-trans map, we analyzed the similarity between the predicted K-trans map generated by deep learning and the ground truth K-trans map derived using experimental data from the DCE protocol. To investigate any advantage of adding a spatial network, we displayed the comparison between ST-Net and the modified fast-eTofts, a purely temporal network (T-Net). Additionally, we performed an evaluation on the GKM derived low dose image to show the improvement of detecting BBB-opening using deep learning.

Noise from the GKM derived and deep learning predicted Ktrans images was first removed using a 3D median filter with local window-size 3x3x3 from Python library-SciPy. The post-processed WB Ktrans maps were then used to visualize and quantify the performance of the algorithms mentioned above (ST-Net, T-Net, GKM derived low dose image) using structural similarity index (SSIM) (1) [29], peak signal-to-noise ratio (PSNR) (2), Pearson correlation coefficient (PCC) (3), concordance correlation coefficient (CCC) [30], area under the curve (AUC) [33], and normalized root MSE (NRMSE) (4) [32] metrics.

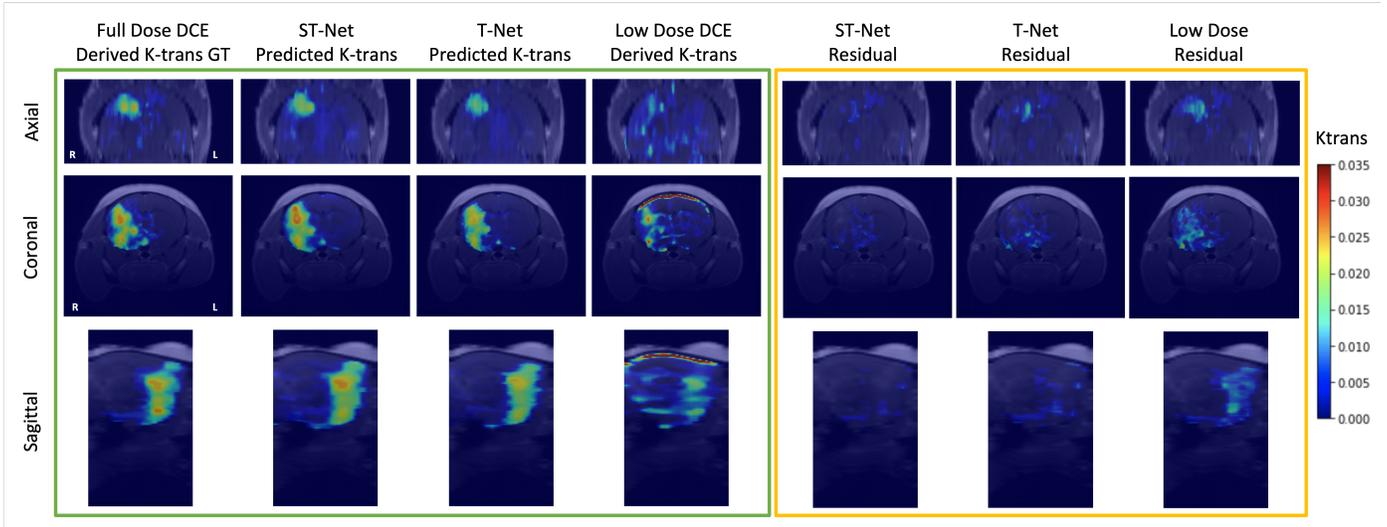

Fig. 3. Full dose and low dose volume transfer constant (Ktrans) map obtained from the general kinetic model (GKM) model and the predicted Ktrans map from two neural networks (green box), along with the residual map between the low-dose/predicted map and full-dose ground truth (orange box).

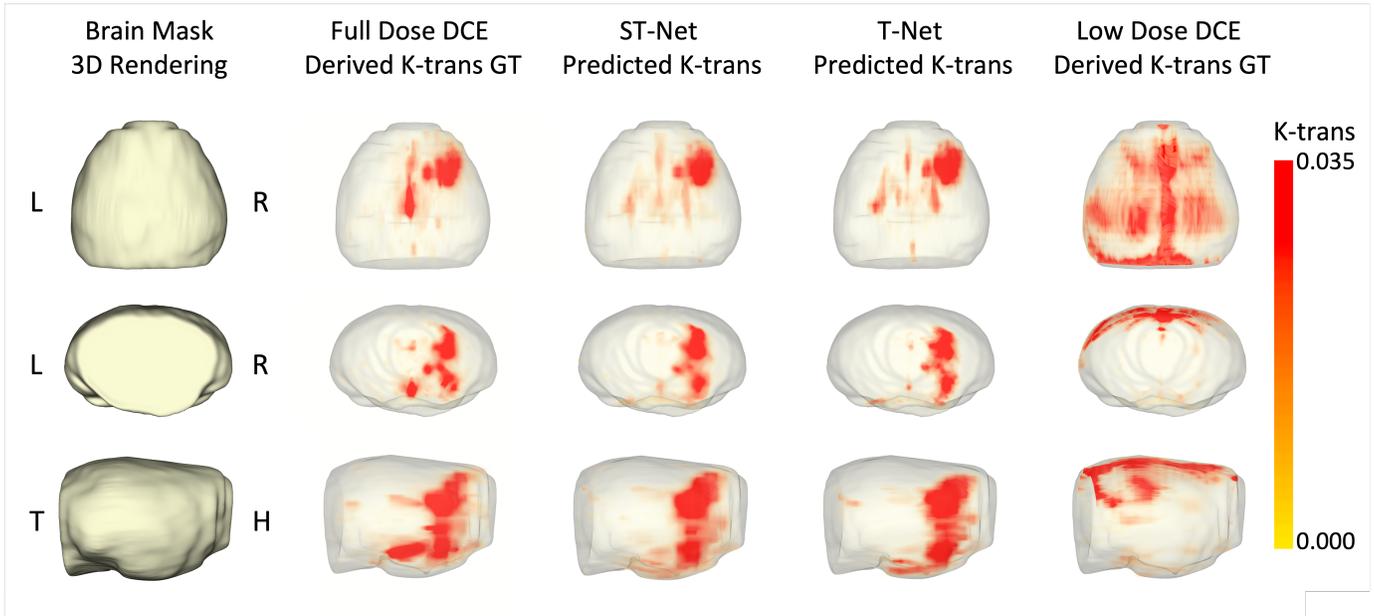

Fig. 4. Mapping the BBB-opening Ktrans map in a three-dimensional brain volume. An "iron" color scheme is applied in the figure. (L: left; R: right; T: tail; H: head)

$$SSIM = l^{\alpha}(x,y)c^{\beta}(x,y)s^{\gamma}(x,y) \quad (1)$$

$$PSNR = 10 \cdot log_{10}(\frac{MAX_x^2}{MSE}) \quad (2)$$

$$PCC = \frac{\sigma_{xy}}{\sigma_x \sigma_y} \quad (3)$$

$$CCC = \frac{2\sigma_{xy}}{(\mu_x - \mu_y)^2 + \sigma_x^2 + \sigma_y^2} \quad (4)$$

$$NRMSE = \frac{\sqrt{\frac{1}{N}\sum(x-y)^2}}{\sqrt{\frac{1}{N}\sum x^2}} \quad (5)$$

Where $x$ and $y$ represent the voxel of ground truth and derived/predicted images. The $l(x,y)$, $c(x,y)$, and $s(x,y)$ in SSIM respectively measure the differences between the luminance, contrast, and structure of the two images, and $\alpha, \beta$, and $\gamma$ are three constants. $MAX_x$ and $MSE$ in PSNR represent the maximum voxel intensity of the ground truth and the mean

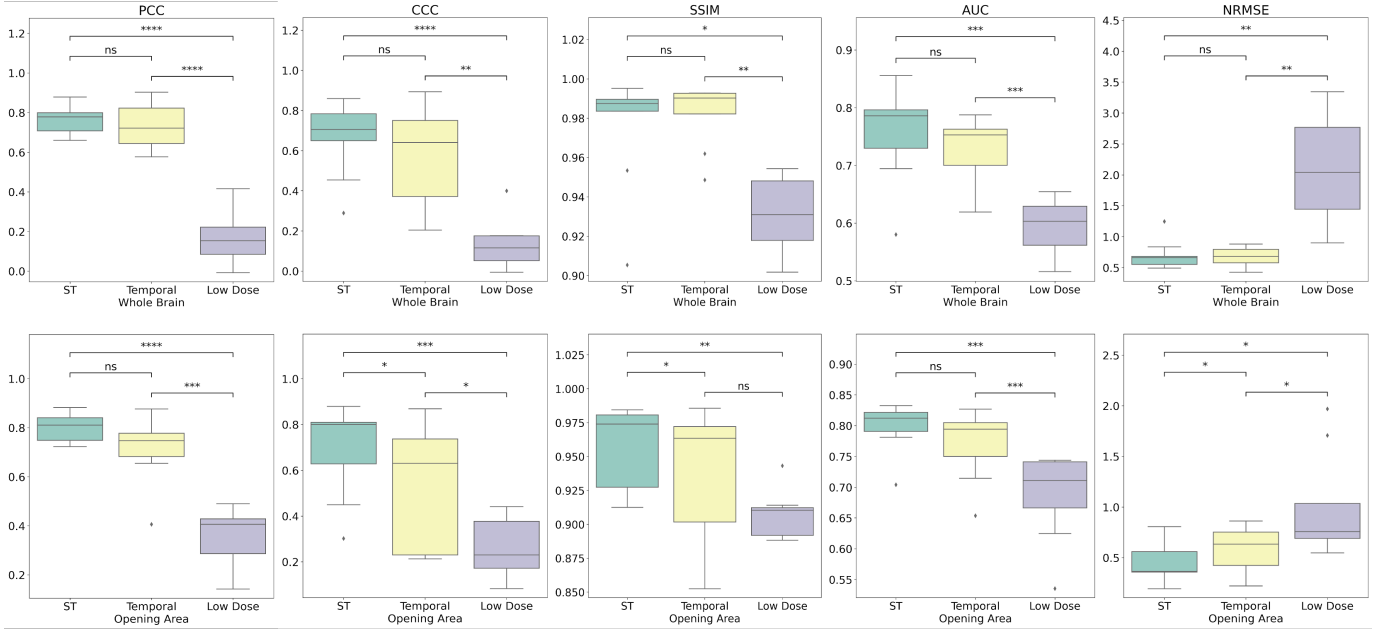

Fig. 5. Box plots visualizing model performance across ten testing mice. Each * indicates order of significance (*: p<0.05; **: p<0.01, ***: p<0.001, ****: p<0.0001).

TABLE I
QUANTITATIVE COMPARISON OF KTRANS MAP OBTAINED FROM GROUND TRUTH AND DERIVED IMAGES FOR ALL SUBJECTS

WB

| Subject | ST | | | | | | Temporal | | | | | | LowDose | | | | | |
|---|---|---|---|---|---|---|---|---|---|---|---|---|---|---|---|---|---|---|
| | SSIM↑ | PSNR↑ | PCC↑ | CCC↑ | AUC↑ | NRMSE↓ | SSIM↑ | PSNR↑ | PCC↑ | CCC↑ | AUC↑ | NRMSE↓ | SSIM↑ | PSNR↑ | PCC↑ | CCC↑ | AUC↑ | NRMSE↓ |
| 1 | 0.990 | 23.634 | 0.878 | 0.860 | 0.834 | 0.493 | 0.991 | 23.751 | 0.863 | 0.806 | 0.753 | 0.488 | 0.902 | 8.485 | 0.204 | 0.116 | 0.603 | 2.768 |
| 2 | 0.905 | 15.624 | 0.790 | 0.454 | 0.786 | 1.246 | 0.982 | 20.858 | 0.823 | 0.538 | 0.788 | 0.683 | 0.943 | 14.292 | 0.154 | 0.145 | 0.641 | 1.445 |
| 3 | 0.986 | 22.440 | 0.855 | 0.832 | 0.796 | 0.522 | 0.993 | 24.279 | 0.903 | 0.893 | 0.780 | 0.425 | 0.931 | 11.996 | 0.222 | 0.177 | 0.629 | 1.728 |
| 4 | 0.989 | 24.301 | 0.799 | 0.784 | 0.789 | 0.593 | 0.990 | 24.398 | 0.766 | 0.750 | 0.752 | 0.590 | 0.922 | 12.353 | 0.085 | 0.057 | 0.587 | 2.341 |
| 5 | 0.991 | 21.625 | 0.708 | 0.705 | 0.783 | 0.682 | 0.990 | 20.950 | 0.648 | 0.640 | 0.763 | 0.736 | 0.918 | 7.701 | 0.106 | 0.052 | 0.605 | 3.343 |
| 6 | 0.987 | 21.067 | 0.660 | 0.649 | 0.694 | 0.684 | 0.993 | 22.525 | 0.721 | 0.708 | 0.700 | 0.578 | 0.948 | 11.194 | -0.008 | -0.006 | 0.516 | 2.042 |
| 7 | 0.997 | 27.078 | 0.693 | 0.676 | 0.908 | 0.716 | 0.996 | 25.487 | 0.557 | 0.511 | 0.865 | 0.858 | 0.956 | 11.684 | 0.060 | 0.026 | 0.629 | 4.137 |
| 8 | 0.953 | 19.749 | 0.671 | 0.289 | 0.580 | 0.836 | 0.949 | 19.298 | 0.644 | 0.204 | 0.619 | 0.881 | 0.954 | 19.142 | 0.302 | 0.176 | 0.562 | 0.900 |
| 9 | 0.984 | 23.106 | 0.778 | 0.723 | 0.730 | 0.552 | 0.962 | 19.976 | 0.643 | 0.308 | 0.673 | 0.797 | 0.954 | 17.312 | 0.416 | 0.399 | 0.654 | 1.080 |
| 10 | 0.995 | 29.090 | 0.760 | 0.654 | 0.856 | 0.659 | 0.993 | 27.234 | 0.576 | 0.372 | 0.753 | 0.818 | 0.914 | 14.996 | 0.081 | 0.039 | 0.537 | 3.313 |
| Avg | 0.978± 0.028 | 22.772 ±3.745 | **0.759± 0.075** | **0.663± 0.174** | **0.775± 0.091** | **0.698± 0.218** | **0.984± 0.016** | **22.876± 2.582** | 0.714± 0.120 | 0.573± 0.227 | 0.745± 0.067 | **0.685± 0.159** | 0.934± 0.019 | 12.916± 3.602 | 0.162± 0.127 | 0.118± 0.118 | 0.596± 0.046 | 2.310± 1.069 |

| Subject | ST | | | | | | Temporal | | | | | | LowDose | | | | | |
|---|---|---|---|---|---|---|---|---|---|---|---|---|---|---|---|---|---|---|
| | SSIM↑ | PSNR↑ | PCC↑ | CCC↑ | AUC↑ | NRMSE↓ | SSIM↑ | PSNR↑ | PCC↑ | CCC↑ | AUC↑ | NRMSE↓ | SSIM↑ | PSNR↑ | PCC↑ | CCC↑ | AUC↑ | NRMSE↓ |
| 1 | 0.983 | 27.269 | 0.841 | 0.835 | 0.832 | 0.357 | 0.971 | 25.195 | 0.839 | 0.730 | 0.827 | 0.429 | 0.888 | 14.958 | 0.286 | 0.172 | 0.711 | 1.707 |
| 2 | 0.944 | 31.899 | 0.856 | 0.799 | 0.833 | 0.317 | 0.902 | 22.630 | 0.747 | 0.313 | 0.794 | 0.649 | 0.888 | 22.502 | 0.393 | 0.234 | 0.742 | 0.690 |
| 3 | 0.981 | 34.305 | 0.882 | 0.879 | 0.822 | 0.193 | 0.986 | 31.535 | 0.876 | 0.869 | 0.805 | 0.220 | 0.910 | 25.158 | 0.428 | 0.377 | 0.741 | 0.547 |
| 4 | 0.974 | 30.246 | 0.810 | 0.810 | 0.821 | 0.365 | 0.963 | 28.435 | 0.771 | 0.737 | 0.809 | 0.423 | 0.943 | 26.088 | 0.490 | 0.441 | 0.744 | 0.641 |
| 5 | 0.977 | 24.394 | 0.728 | 0.697 | 0.781 | 0.560 | 0.972 | 23.060 | 0.683 | 0.631 | 0.767 | 0.633 | 0.909 | 17.047 | 0.142 | 0.083 | 0.666 | 1.969 |
| 6 | 0.984 | 30.026 | 0.817 | 0.806 | 0.812 | 0.363 | 0.983 | 27.691 | 0.777 | 0.737 | 0.801 | 0.406 | 0.912 | 22.426 | 0.153 | 0.119 | 0.669 | 1.035 |
| 7 | 0.984 | 35.007 | 0.813 | 0.768 | 0.843 | 0.494 | 0.973 | 32.256 | 0.701 | 0.600 | 0.812 | 0.607 | 0.938 | 32.923 | 0.419 | 0.416 | 0.729 | 0.901 |
| 8 | 0.912 | 21.067 | 0.723 | 0.301 | 0.704 | 0.806 | 0.902 | 20.507 | 0.695 | 0.213 | 0.715 | 0.862 | 0.914 | 20.514 | 0.438 | 0.192 | 0.625 | 0.857 |
| 9 | 0.922 | 27.249 | 0.748 | 0.628 | 0.794 | 0.494 | 0.853 | 23.059 | 0.655 | 0.221 | 0.750 | 0.767 | 0.910 | 24.706 | 0.425 | 0.389 | 0.741 | 0.757 |
| 10 | 0.927 | 26.440 | 0.770 | 0.449 | 0.791 | 0.615 | 0.892 | 23.097 | 0.406 | 0.230 | 0.654 | 0.752 | 0.892 | 20.796 | 0.406 | 0.230 | 0.535 | 0.752 |
| Avg | **0.959± 0.029** | **28.790 ±4.366** | **0.799± 0.055** | **0.697± 0.187** | **0.803± 0.040** | **0.456± 0.175** | 0.940± 0.048 | 25.747± 4.025 | 0.715± 0.129 | 0.528± 0.256 | 0.773± 0.054 | 0.575± 0.200 | 0.911± 0.019 | 22.712± 5.016 | 0.358± 0.122 | 0.265± 0.130 | 0.690± 0.068 | 0.985± 0.474 |

(PCC: Pearson correlation coefficient; CCC: concordance correlation coefficient; SSIM: structural similarity index; AUC: area under the ROC curve; NRMSE: normalized root mean square error) square error of the two images. $\mu_x$ and $\mu_x$ are the means for the two images, and $\sigma_x$ and $\sigma_y$ are the corresponding variances.

$\sigma_{xy}$ is the covariance and N is the voxel number within the ROI. Both ST-Net and T-Net networks were trained using five-fold cross-validation and Student's t-test was performed on the metrics mentioned above. Significant differences are shown in box plots, with each number of * indicating the order of significance (*: p<0.05; **: p<0.01, ***: p<0.001, ****: p<0.0001).

### G. Animal Model

We used ten C576J/BL mice at the age of 3-6 months old for this study. Mice were scanned using the protocol proposed in the DCE-MRI protocol described previously. A total of 162 scans were acquired for ten subjects.

## III. RESULTS

We compared our proposed ST-Net with low contrast agent dosage Ktrans images derived from the conventional GKM model and the temporal-only deep learning model, T-Net. The derived/predicted 2D Ktrans images for one testing subject from three different orientations were visualized in Fig. 3. As shown in Fig. 3, the first column was the full dose Ktrans images derived by conventional GKM fitting and was used as ground truth in the deep learning model. The second to fourth columns were low-dose Ktrans images mapped by the GKM model, predictions by T-Net, and predictions by ST-Net, respectively. The following four columns displayed the residual differences between full dose and derived/predicted Ktrans images. Additionally, three-dimensional rendering of BBB-opening for low dose, full dose, T-Net, and ST-Net are shown in Fig. 4.

The quantitative comparison among low dose, T-Net, and ST-Net on all the ten testing subjects for WB and opening area, are summarized in Table 1. The average performance with standard deviation is shown in the last row of the Table. In the post-processed reconstructed WB Ktrans maps, ST-Net achieved the highest PCC (0.759 ± 0.075), CCC (0.663 ± 0.174), and AUC (0.775 ± 0.091) across ten testing mice. On the other hand, T-Net performed better in SSIM (0.984 ± 0.016), PSNR (22.876 ± 2.582), and NEMSE (0.685 ± 0.159). For the BBB-opening area, the ST-Net model outperformed in every metrics (SSIM = 0.959 ± 0.029, PSNR = 28.790 ± 4.366, PCC = 0.799 ± 0.055, CCC = 0.697 ± 0.187, AUC = 0.803 ± 0.040, and NRMSE = 0.456 ± 0.175).

The box plots across the ten testing mice for each metric in two ROIs are shown in Fig. 5. Fig. 5 also shows that both T-Net and ST-Net have significant differences compared to low dose Ktrans images in both WB and BBB-opening only areas. ST-Net and T-Net also show significant differences in the opening areas for every metric.

## IV. Discussion

FUS with intravenous administration of microbubbles has been shown to open the BBB in small animals in-vivo and in clinical trials. This opening can be targeted and is transient. The FUS-enhanced BBB-opening can be validated with MRI, Positron emission tomography (PET), Single-photon emission computed tomography (SPECT), etc., using labeled molecules. Given the limited tolerance of GBCA, novel approaches are needed for clinical applicability [44]. In this study, we proposed ST-Net, a spatiotemporal CNN deep learning architecture designed for predicting a full dose time-series BBB-opening by low dose T1W MRI. We not only successfully investigated the efficacy of detecting BBB-opening with low dosage contrast agent administration but also improved the model performance with an additional 3D CNN.

We first validated a deep learning algorithm that can be used to acquire full dose Ktrans maps while decreasing contrast agent dosage using T-Net. The comparison from three directions in Fig. 3 shows a high similarity between T-Net and ground truth. Compared to low dose derived K-trans, T-Net depicts the BBB-opening area and outlines more accurately. However, we noticed the edges of the BBB-opening in T-Net look noisy in the residual maps. The reason is that we only focus on temporal information in T-Net, therefore, the model cannot differentiate the boundaries between opening and non-opening tissues. The intensity observed within the BBB-opening being lower than the ground truth was caused by the same reason. The intensity of the FUS focus point of the BBB-opening in the Ktrans map should be the highest. However, since T-Net only learned the contrast agent concentration changes for each pixel, it cannot detect the intensity difference among adjacent pixels.

As a result, we proposed adding a spatial network to share the features across the brain.

One of our main contributions is the novelty of adding a spatial network to further enhance the performance of predicting Ktrans while retaining high fidelity. Instead of simply inputting data on the voxel level, we cropped the WB ROI to patches across time and extracts the spatial features for each patch through a three-dimensional CNN encoder. With spatial information, ST-Net was able to learn the brain structure, and predict the BBB-opening location and shape in reference to the neighbor voxels. The t-test results in Fig. 5 shows there are significant differences between ST-Net predicted results and GKM derived low dose BBB-opening among all metrics in both ROIs. The 2D whole-head (WH) Ktrans images overlapped with structural MRI scans in Fig. 3 and the 3D WB volumes in Fig. 4 visualize one of the testing subjects. Both figures show a clear BBB-opening in ST-Net; however, we can barely visualize the opening in the low dose image. The results demonstrate the efficiency and potential of using ST-Net with a low-dose contrast agent in detecting BBB-opening.

The advantages of adding a spatial network in ST-Net include increasing model robustness and improving the prediction at the edges of the BBB-opening. The box plot in Fig. 5 shows the SSIM, PSNR, CCC, and SSIM significantly increase and the NRMSE significantly decreases within the opening area in ST-Net. The standard deviations of ST-Net for all metrics for both ROIs are the smallest, demonstrating that the spatial network was a crucial element in predicting 3D images by providing spatial information from neighbor voxels. The PSNR of the ST-Net opening area is significantly improved in Fig. 5 shows ST-Net provides a denoise effect. The statistical improvement can be visualized in the sagittal direction in Fig. 3. The comparison shows that ST-Net is predicted better on the opening boundary and the opening edge was less noisy than T-Net model. Moreover, the intensity of the BBB-opening in ST-Net was observed to be more similar to the ground truth and matches the structure of the BBB-opening. As a result, adding a spatial network proved to be of high value/importance.

ST-Net not only predicted the BBB-opening area accurately, but the normal-appearing brain tissues also show a high resemblance to the ground truth. The BBB-opening might not be induced by FUS in some cases; therefore, the ability of modeling non-opening areas is as critical as modeling opening areas. Moreover, in practice, the BBB-opening will not be limited to one position, therefore, the BBB-opening locations were different across ten datasets in our data to increase data diversity and to simulate the realistic clinical application. The non-opened areas were used for negative control. The 2D WH Krans images overlapped with structural MRI scans and the 3D brain volumes for a subject without BBB-opening as shown in the supplementary material.

Some future directions may further improve the utility of ST-Net. First, the FUS parameters we used in the experiment were consistent, therefore the range of the BBB-opening was not much different across subjects. Additionally, although we collected BBB-opening data for both sides of the brain, the locations of the BBB-opening were restricted within the striatum region. Finally, we also can discuss the MRI images acquired with low-field scanners [40] or state-of-the-art portable or mobile MRI machines [41] to acquire images with less scanning time or offer to extend beyond the traditional hospital and imaging center boundaries. To apply our method in a clinical setting, we should replicate and validate the entire experiment with a variety of samples such as conducting high-pressure FUS treatments, collecting BBB-opening data in distinct brain areas, and/or utilizing different types of MRI scanners or MR systems.

This study was also limited by the sample diversity. In this study, we only investigated the efficacy of detecting BBB-opening with low dosage contrast agent administration in healthy wild-type mice brains. However, the long term goal of this research is to apply the method to those who suffer from CNS disease. As a result, validating our results on brain tumor or Alzheimer's disease mouse models is necessary. Nonetheless, to extend the research to clinical trials, future non-human primate studies should confirm the result reported in mice studied here.

Although the current model is highly automatic, there are some details in the preprocessing pipeline we can improve. The entire deep learning model has been designed to be fully automatic if the DCE-MRI scans and ROI masks are ready. However, we manually labeled the WB and brain muscle regions for the preprocessing step. Even though it is much easier to label muscle maps compared to the reference paper labeling blood vessels, the annotation process is still time-consuming and can disagree depending on the researcher. A possible resolution is to train an additional deep learning-based segmentation tool as described in [36] to remove skull artifacts and extract ROI through a simple U-Net architecture.

There are several strategies we can do to further improve the performance or efficiency of the model. In ST-Net, we train the spatial information first and then fuse it with the temporal features. We could try to switch the order sequence of the spatial and temporal network to see if there is any improvement. Furthermore, instead of training spatial and temporal models in sequence, we can design our spatiotemporal model to two-stream CNN, training spatial and temporal models in parallel, and combining the two networks with a late fusion technique. Another strategy is to substitute CNN with different networks such as LSTM [37] or Transformer. In addition to revising the model architecture, we can also implement the data augmentation technique and transfer learning to retrain the model.

Even though we successfully estimated BBB-opening with reduced GBCA, GBCA is still necessary. GBCAs can lead to severe side effects in some patients, especially those suffering from kidney disease. Therefore, it is crucial to eliminate the usage of contrast agents in clinical research by developing another framework to achieve a contrast-free method. In ST-Net, we use DCE-MRI and T1W scans as input. In future work, we can introduce multi-modality non-contrast MRI sequences such as effective T2 (T2*), susceptibility-weighted (SWI), arterial spin labeling (ASL), and diffusion tensor (DTI).

## V. CONCLUSION

In conclusion, we validated the hypothesis of implementing a neural network model focusing on the temporal domain with time-series DCE-MRI data to model Ktrans and detect BBB-opening with low-dose GBCAs. Furthermore, we added a spatial CNN network to significantly improve the performance of Ktrans-based BBB opening confirmation performance. We showed the potential of reducing the use of GBCAs and reducing the risk of contrast agent-induced side effects, thereby improving the safety profile of FUS treatments in the brain. Our data is publicly available, and our code can be found on GitHub.

## SUPPLEMENTAL

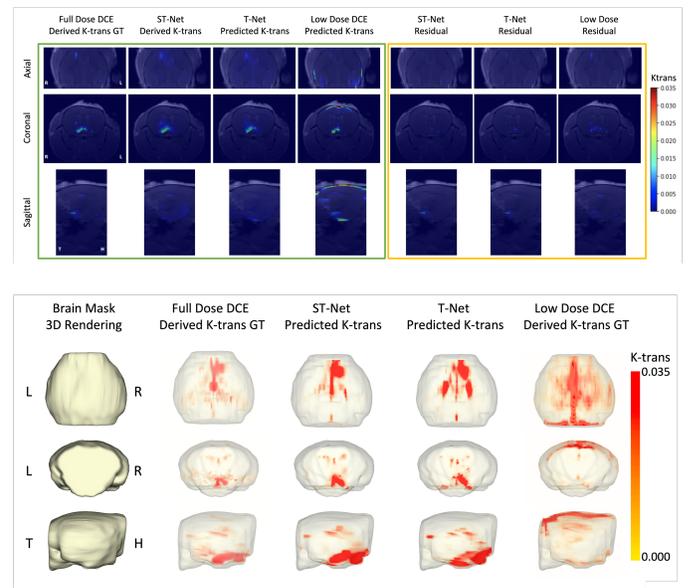

Fig. 6. The 2D WH Krans images overlapped with structural MRI scans and the 3D brain volumes for an example subject without BBB-opening. (L: left; R: right; T: tail; H: head)

## ACKNOWLEDGMENT

This research was funded by the Gary and Yael Fegel Family Foundation, St. Baldrick's Foundation, the Star and Storm Foundation, the Matheson Foundation (UR010590), Swim Across America, a Herbert Irving Cancer Center Support Grant (P30CA013696), Sebastian Strong Foundation, National Institutes of Health Grants (5R01EB009041) and (5R01AG038961), and the ZI Seed Grant for MR Studies Program. The content is solely the responsibility of the authors and does not necessarily represent the official views of the NIH.